# Manipulation of the magnetic configuration of (Ga,Mn)As nanostructures

J.A. Haigh, M. Wang, A.W. Rushforth, E. Ahmad, K.W. Edmonds, R.P. Campion, C.T. Foxon, and B.L. Gallagher

School of Physics and Astronomy, University of Nottingham, University Park, Nottingham, NG7 2RD, United Kingdom

*We have studied the magnetic reversal of L-shaped nanostructures fabricated from (Ga,Mn)As. The strain relaxation due to the lithographic patterning results in each arm having a uniaxial magnetic anisotropy. Our analysis confirms that the magnetic reversal takes place via a combination of coherent rotation and domain wall propagation with the domain wall positioned at the corner of the device at intermediate stages of the magnetic hysteresis loops. The domain wall energy can be extracted from our analysis. Such devices have found implementation in studies of current induced domain wall motion and have the potential for application as non-volatile memory elements.*

75.50.Pp, 75.30.Gw, 75.60.Jk, 75.60.-d, 75.75.+a

The micromagnetic behavior of magnetic nanowires is a topic of widespread interest, due to the potential of such structures for new non-volatile memory and storage applications[1]. The dilute magnetic semiconductor (Ga,Mn)As exhibits many properties which make it well-suited for the study of micro-magnetic and magneto-transport phenomena. The strong spin-orbit coupling of the valence band holes which mediate the ferromagnetic interaction between the substitutional $Mn^{2+}$ ions can give rise to large, strain dependent magneto-crystalline anisotropies[2] and anomalous magneto-transport coefficients[3]. This makes it possible to induce strongly in-plane or out-of-plane magnetic anisotropy by growth induced strain, due to lattice matching to the substrate[2]. Recently control of anisotropy has been achieved by both local lattice relaxation in etched nanoscale structures[4,5] and electrically using a piezoelectric transducer[6]. In addition to the ability to engineer the magnetic anisotropy, (Ga,Mn)As is also a low moment system with large magnetic stiffness resulting in single domain characteristics. These favorable properties have been utilized in recent studies of current induced domain wall motion by spin transfer torque[5] and in nanoscale structures exhibiting large magnetoresistance with the potential for application as non-volatile memory[7]. Both of these studies involved measuring the electrical transport properties of nanoscale devices with an L-shaped geometry and the results were interpreted as being due to the switching of the magnetization of each arm of the device separately with the formation of a magnetic domain wall at the corner of the device. However, in these studies the magnetic configuration was not directly demonstrated but inferred from electrical measurements.

In this paper we present a magnetometry study of nanoscale (Ga,Mn)As L-shaped devices. Our analysis confirms that the observed magnetic hysteresis is consistent with each arm of the device behaving as a single domain and magnetic reversal involving the formation of a domain wall at the corner of the device. We are able to extract the domain wall energy, which is the same order of magnitude as the uniaxial anisotropy of each arm of the device, consistent with the notion that magnetic reversal takes place via the propagation of a 180º domain wall.

L-shape devices with arms of length 20μm and width 750nm oriented along the [110] and $[1\bar{1}0]$ directions were fabricated by electron beam lithography and wet chemical etching of a 25nm $Ga_{0.94}Mn_{0.06}As$ layer grown by low temperature molecular beam epitaxy on a GaAs(001) substrate and buffer layers[8]. Approximately $10^5$ devices, separated by 1.5μm were fabricated on a 5mm x 5mm chip. Figure 1 shows a scanning electron beam image of an area of the device array. Magnetization measurements were carried out using a Quantum Design MPMS SQUID magnetometry system.

The unpatterned (Ga,Mn)As layer shows the usually observed magnetic behavior. It has a hard magnetic axis perpendicular to plane due to the compressive growth strain[2] and the in-plane magnetic easy axis is determined by a competition between a cubic [100]/[010] anisotropy and a uniaxial anisotropy favoring the $[1\bar{1}0]$ direction[9]. For this wafer the uniaxial anisotropy is the dominant term resulting in the easy axis along the $[1\bar{1}0]$ direction for temperatures down to T=2K. For lithographically defined sub-micron bars of (Ga,Mn)As, local strain relaxation at the edges introduces a further uniaxial magnetic anisotropy[4]. The magnetic free energy of such bars can be expressed as

$$E = K_{U,i}\sin^2(\theta+45) + (K_C/4)\sin^2(2\theta) - MH\cos(\varphi-\theta) \qquad (1)$$

where $K_C$ is the cubic anisotropy constant, and the uniaxial anisotropy constant $K_{U,i}$ can have contributions from both the "intrinsic" uniaxial anisotropy and that due to patterning the bar along the i=[110] or [1$\bar{1}$0] directions. M is the saturation magnetization at angle $\theta$ to the [100] direction and H is the external magnetic field at angle $\varphi$ to the [100] direction.

We model the L-shape devices assuming that the magnetization of each arm rotates and switches independently of the other. M-H hysteresis curves are obtained by minimizing the energy density in equation (1), with the additional condition that magnetization switches between local energy minima occur when the reduction in magnetic energy density in equation (1) equals the increase in energy density, $\varepsilon$ due to forming a domain wall. To allow for a statistical variation of $\varepsilon$ across the $10^5$ devices a Gaussian distribution with mean value $\varepsilon_d$ and standard deviation $\sigma$ was used.

Figures 2 (a)-(c) show the magnetization of the L-shaped bars measured at 2K for external magnetic fields applied along the [110], [1$\bar{1}$0] and [010] directions and the best fits obtained using the model detailed above. The same anisotropy and domain wall energies are used in the fitting for the three applied field directions. In units of Jm$^{-3}$ these are $K_C = 310\pm10$; $K_{U,[110]} = 380\pm30$; $K_{U,[1\bar{1}0]} = -950\pm40$; $\varepsilon_d = 480$; and $\sigma = 104$. The same procedure was used to fit to the hard axis hysteresis loop of an unpatterned control sample yielding $K_C=210\pm20$ and $K_U=-330\pm30$.

The quality of this fitting is excellent, and an analysis of the results provides a consistent picture of the magnetization reversal processes in the device array. The remanence as a fraction of the saturation magnetization along the [110] and [1$\bar{1}$0] directions is close to ½ while that along [010] is close to $1/\sqrt{2}$ showing that the remanent magnetizations are along the long axes of the bars. This is due to the dominance of the uniaxial anisotropy induced by the lithographic patterning, which is also revealed by comparing the anisotropy constants for the patterned and unpatterned samples. The lithographic patterning also induces a small change in $K_C$. From symmetry considerations a uniaxial symmetry breaking can be expected to lead to changes in the higher order anisotropy constants, as it does in the transport coefficients[3], however the effect on the first order uniaxial anisotropy constant is clearly more significant. For the external field applied along the [110] and [1$\bar{1}$0] directions (Figs 2(a) and (b)) the magnetization reversal consists of coherent rotation towards the hard axis for the arm perpendicular to the field, while for the arm parallel to the field the reversal involves the nucleation of a domain wall. The reversal of the arm parallel to the field occurs first, resulting in the domain wall being positioned at the corner of the L-shape device. For the field applied along the [010] direction the reversal of each arm includes a combination of rotation and switching with the formation of a domain wall which is positioned at the corner of the device.

The domain wall energy density obtained is $\varepsilon_d = 480$ Jm$^{-3}$ ($\sigma = 104$ Jm$^{-3}$). The predicted domain wall energy density is $\sim \pi(AK)^{1/2}/d$, with A$\sim$0.1pJ/m[10,11], and the typical domain wall thickness is $d\sim$50nm for (Ga,Mn)As films with in-plane easy magnetic axes[11]. The extracted value of $\varepsilon_d$ therefore gives an effective anisotropy energy $K \sim 580$ J/m$^3$. This is similar to the uniaxial anisotropy energies of the individual arms and thus consistent with reversal involving the propagation of a near 180º domain wall along the bar.

A more detailed analysis of the hysteresis loops in Figs 2 (b) and (c) reveals that for the arms patterned along the [110] direction the reversal involves a rapid rotation of the magnetization for a small applied magnetic field. These transitions are labeled (i) and (ii) in Figs. 2 (b) and (c) respectively. Similar transitions are not observed for the arms patterned along the [1$\bar{1}$0] direction. The transitions are due to the interplay of cubic and uniaxial anisotropies of similar magnitude, causing the position of the energy minima defined by equation 1 to become rather susceptible to a small applied field [12]. The patterning induced anisotropy has the same sign as the "intrinsic" uniaxial anisotropy for the [1$\bar{1}$0] arms resulting in very strong [1$\bar{1}$0] uniaxial anisotropy while the two uniaxial contributions have opposite sign for the [110] arms giving a net [110] uniaxial anisotropy comparable in magnitude to the cubic anisotropy at low temperature. Figure 3 shows that the small angle biaxial transition in the [110] arm becomes less distinct as temperature is increased and cannot be resolved above 30K. This is because the cubic anisotropy energy decreases more rapidly than the uniaxial anisotropy as temperature increases[9] and both arms become dominated by the uniaxial anisotropy.

Micropatterned bars of (Ga,Mn)As, with similar geometry to those in our samples, were investigated previously[5]. The micromagnetic behavior, inferred from magnetotransport measurements, showed evidence for magnetization switching driven by spin transfer torque. The present results give a direct confirmation of the domain patterns inferred in such measurements, showing single-domain-like characteristics over each arm of the devices, with reversal by propagation of near-180$^o$ walls. They also show a remarkable uniformity of the magnetic properties across the ~$10^5$ devices making up the array. However, the magnetic signal from the corner of the devices, where trapping of the domain walls occurs, makes up only a small fraction of the total signal, so that further studies are needed to determine the detailed magnetic domain structure in this region.

In summary, we have studied the magnetic reversal of L-shaped (Ga,Mn)As nanostructures. The strain relaxation due to the lithographic patterning results in a uniaxial magnetic anisotropy along each arm. Our analysis confirms that the magnetic reversal takes place via a combination of coherent rotation and domain wall propagation with the domain wall being position at the corner of the device at intermediate stages of the magnetic hysteresis loops. Such devices are useful in studies of current induced domain wall motion and have the potential for application as non-volatile memory elements.

We acknowledge funding from EU grant NAMASTE 214499 and EPSRC grant GR/S81407/01.

**Figure 1** Scanning electron microscope image of the L-shaped bars fabricated from the (Ga,Mn)As wafer.

**Figure 2 (Color online)** Magnetic hysteresis loops measured at 2K along the (a) [110], (b) [1$\bar{1}$0] and (c) [010] directions. Black squares show the measured data. Red circles show the fitted curves. The cartoons show the magnetic reversal of each arm at various points along the curve for field sweeping up.

**Figure 3 (Color online)** Magnetic hysteresis loops for the field applied along the [1$\bar{1}$0] direction for a range of temperatures.

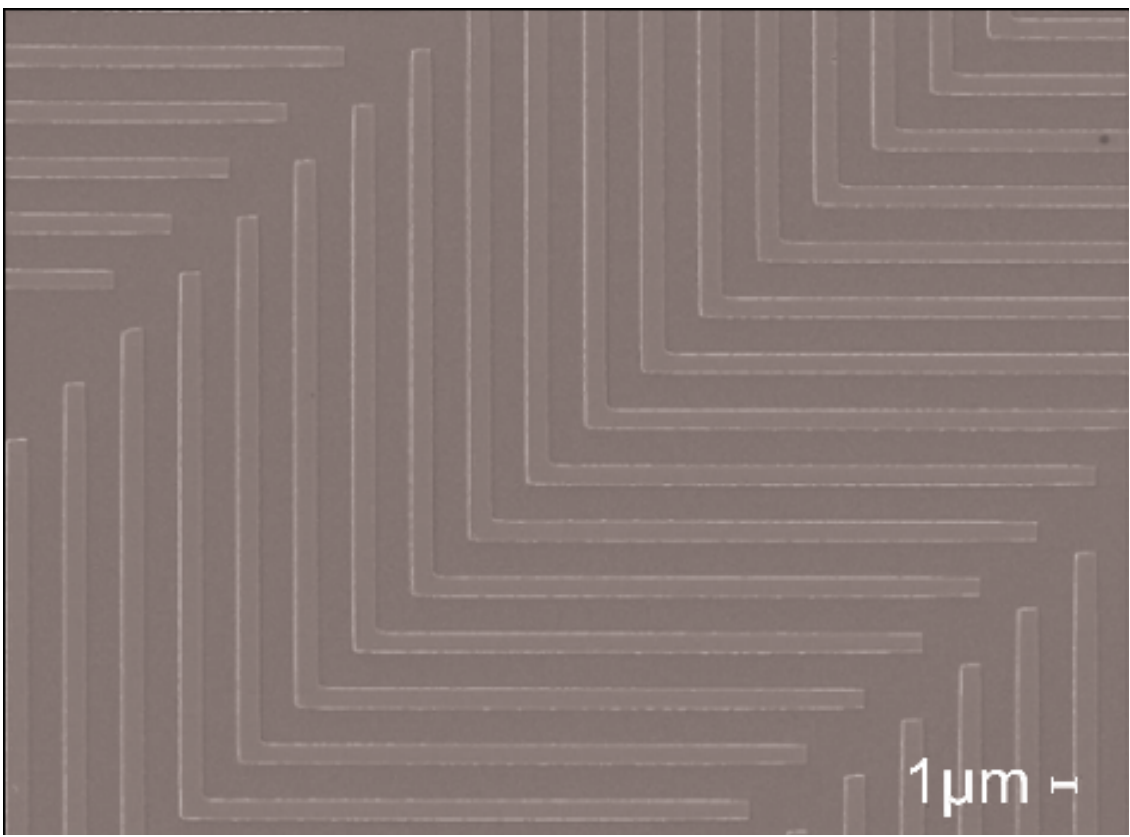

FIGURE 1

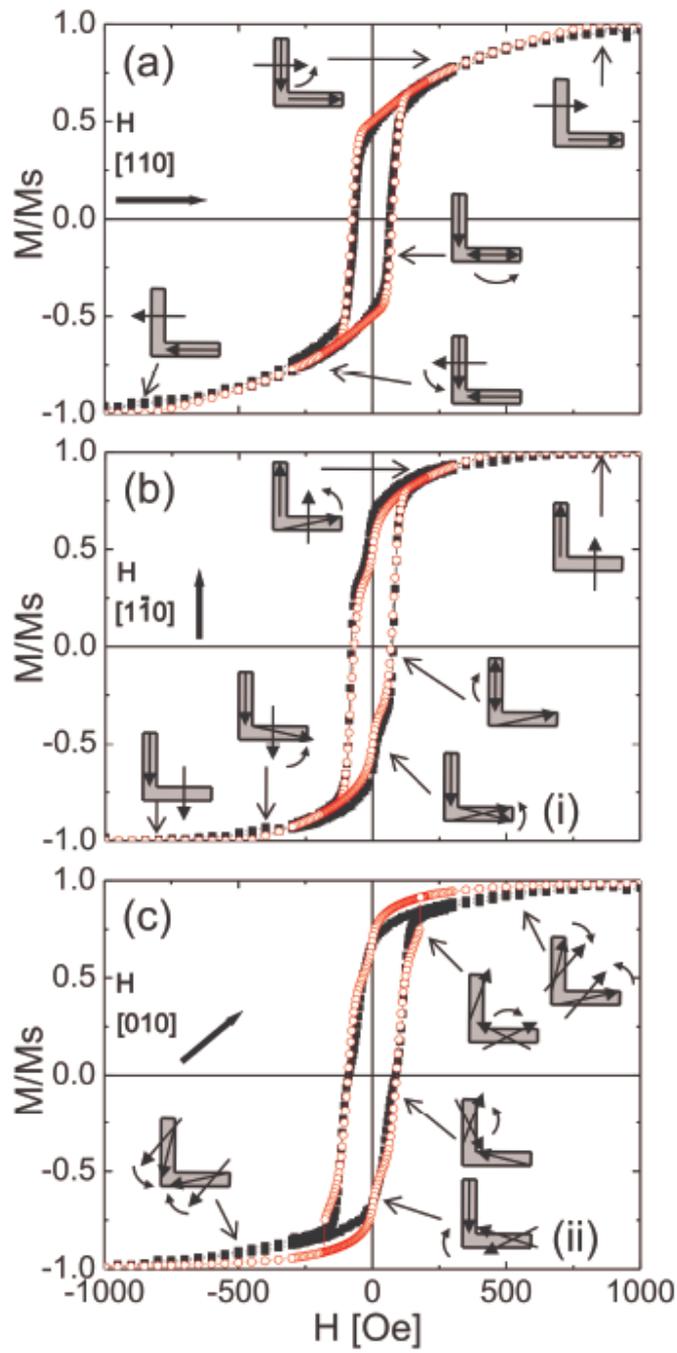

FIGURE 2

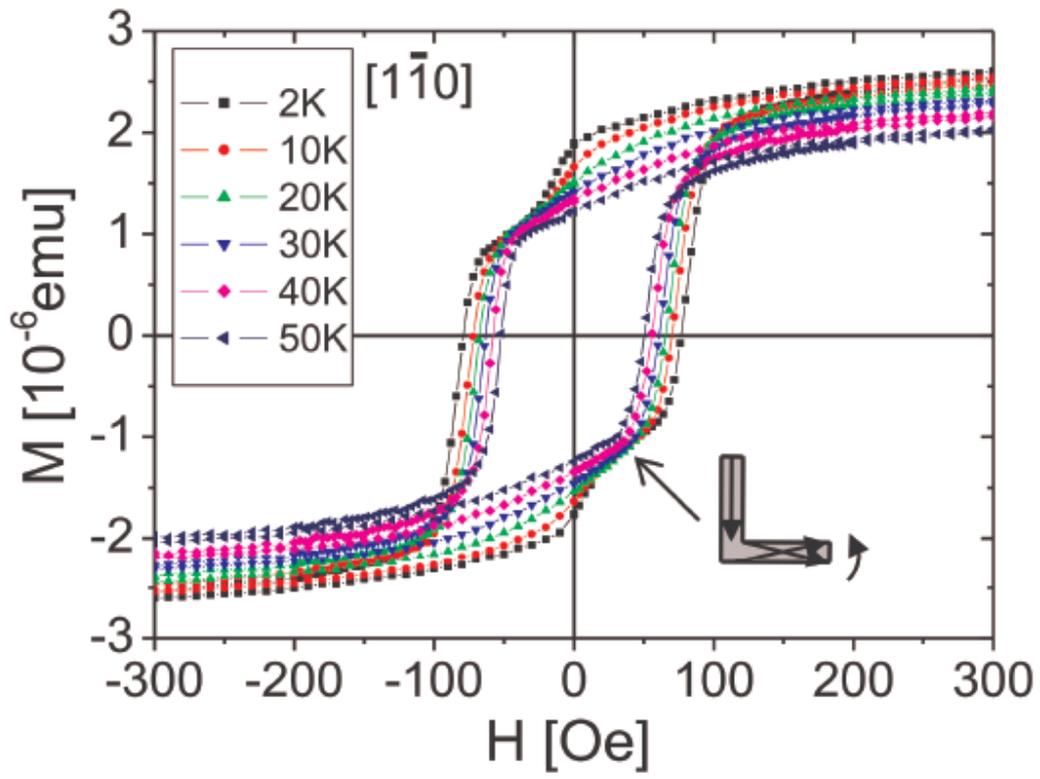

FIGURE 3